\documentclass[12pt]{article}
\newcommand{\nc}{\newcommand}
\nc{\renc}{\renewcommand}

\usepackage{graphicx}
\nc{\half}{{\textstyle{1\over2}}}
\nc{\etal}{\mbox{\it et al. }}
\nc{\ie}{{\it i.e.}}
\nc{\eg}{{\it e.g.}}

\renc{\thefootnote}{\arabic{footnote}}
\nc{\capt}[1]{{\bf Figure.} {\small\sl #1}}

\nc{\eqs}[2]{\mbox{Eqs.~(\ref{#1},\,\ref{#2})}}
\nc{\eq}[1]{\mbox{Eq.~(\ref{#1})}}

\nc{\figs}[2]{\mbox{Figs.~(\ref{#1},\,\ref{#2})}}
\nc{\fig}[1]{\mbox{Fig~.(\ref{#1})}}

\nc{\tag}[1]{\label{#1} \marginpar{{\footnotesize #1}}}
\nc{\mtag}[1]{\label{#1} \mbox{\marginpar{{\footnotesize #1}}}}
\renc{\baselinestretch}{1.5}
\newlength{\overeqskip}
\newlength{\undereqskip}
\nc{\be}[1]{\begin{equation} \mbox{$\label{#1}$}}
\nc{\bea}[1]{\begin{eqnarray} \mbox{$\label{#1}$}}
\nc{\Section}[2]{\section{#2}\label{#1}}
\nc{\Bibitem}[1]{\bibitem{#1}}
\nc{\Label}[1]{\label{#1}}
\nc{\eea}{\vspace{\undereqskip}\end{eqnarray}}
\nc{\ee}{\vspace{\undereqskip}\end{equation}}
\nc{\bdm}{\begin{displaymath}}
\nc{\edm}{\end{displaymath}}
\nc{\dpsty}{\displaystyle}
\nc{\bc}{\begin{center}}
\nc{\ec}{\end{center}}
\nc{\ba}{\begin{array}}
\nc{\ea}{\end{array}}
\nc{\bab}{\begin{abstract}}
\nc{\eab}{\end{abstract}}
\nc{\btab}{\begin{tabular}}
\nc{\etab}{\end{tabular}}
\nc{\bit}{\begin{itemize}}
\nc{\eit}{\end{itemize}}
\nc{\ben}{\begin{enumerate}}
\nc{\een}{\end{enumerate}}
\nc{\bfig}{\begin{figure}}
\nc{\efig}{\end{figure}}
\nc{\arreq}{&\!=\!&}
\nc{\arrmi}{&\!-\!&}
\nc{\arrpl}{&\!+\!&}
\nc{\arrap}{&\!\!\!\approx\!\!\!&}
\nc{\non}{\nonumber\\*}
\nc{\align}{\!\!\!\!\!\!\!\!&&}

\def\lsim{\; \raise0.3ex\hbox{$<$\kern-0.75em
      \raise-1.1ex\hbox{$\sim$}}\; }
\def\gsim{\; \raise0.3ex\hbox{$>$\kern-0.75em
      \raise-1.1ex\hbox{$\sim$}}\; }
\nc{\DOT}{\hspace{-0.08in}{\bf .}\hspace{0.1in}}
\nc{\Laada}{\hbox {$\sqcap$ \kern -1em $\sqcup$}}
\nc\loota{{\scriptstyle\sqcap\kern-0.55em\hbox{$\scriptstyle\sqcup$}}}
\nc\Loota{{\sqcap\kern-0.65em\hbox{$\sqcup$}}}
\nc\laada{\Loota}
\nc{\qed}{\hskip 3em \hbox{\BOX} \vskip 2ex}

\nc{\real}{{\rm I \! R}}
\nc{\Z}{{\sf Z \!\!\! Z}}
\nc{\complex}{{\rm C\!\!\! {\sf I}\,\,}}
\def\bigid{\leavevmode\hbox{\small1\kern-3.8pt\normalsize1}}
\def\id{\leavevmode\hbox{\small1\kern-3.3pt\normalsize1}}
\nc{\slask}{\!\!\!/}
\nc{\bis}{{\prime\prime}}
\nc{\pa}{\partial}
\nc{\na}{\nabla}
\nc{\ra}{\rangle}
\nc{\la}{\langle}
\nc{\goto}{\rightarrow}
\nc{\swap}{\leftrightarrow}

\nc{\EE}[1]{ \mbox{$\cdot10^{#1}$} }
\nc{\abs}[1]{\left|#1\right|}
\nc{\at}[2]{\left.#1\right|_{#2}}
\nc{\norm}[1]{\|#1\|}
\nc{\abscut}[2]{\Abs{#1}_{\scriptscriptstyle#2}}
\nc{\vek}[1]{{\rm\bf #1}}
\nc{\integral}[2]{\int\limits_{#1}^{#2}}
\nc{\inv}[1]{\frac{1}{#1}}
\nc{\dd}[2]{{{\partial #1}\over{\partial #2}}}
\nc{\ddd}[2]{{{{\partial}^2 #1}\over{\partial {#2}^2}}}
\nc{\dddd}[3]{{{{\partial}^2 #1}\over
        {\partial #2 \partial #3}}}
\nc{\dder}[2]{{{d #1}\over{d #2}}}
\nc{\ddder}[2]{{{d^2 #1}\over{d {#2}^2}}}
\nc{\dddder}[3]{{d^2 #1}\over
        {d #2 d #3}}
\nc{\dx}[1]{d\,^{#1}x}
\nc{\dy}[1]{d\,^{#1}y}
\nc{\dz}[1]{d\,^{#1}z}
\nc{\dl}[1]{\frac{d\,^{#1}l}{(2\pi)^{#1}}}
\nc{\dk}[1]{\frac{d\,^{#1}k}{(2\pi)^{#1}}}
\nc{\dq}[1]{\frac{d\,^{#1}q}{(2\pi)^{#1}}}

\nc{\cc}{\mbox{$c.c.$ }}
\nc{\hc}{\mbox{$h.c.$ }}
\nc{\cf}{cf.\ }
\nc{\erfc}{{\rm erfc}}
\nc{\Tr}{{\rm Tr\,}}
\nc{\tr}{{\rm tr\,}}
\nc{\pol}{{\rm pol}}
\nc{\sign}{{\rm sign}}
\nc{\bfT}{{\bf T }}

\def\GeV{{\rm\ GeV}}
\def\MeV{{\rm\ MeV}}

\nc{\cA}{{\cal A}}
\nc{\cB}{{\cal B}}
\nc{\cD}{{\cal D}}
\nc{\cE}{{\cal E}}
\nc{\cG}{{\cal G}}
\nc{\cH}{{\cal H}}
\nc{\cL}{{\cal L}}
\nc{\cO}{{\cal O}}
\nc{\cT}{{\cal T}}
\nc{\cN}{{\cal N}}
\nc{\rvac}[1]{|{\cal O}#1\rangle}
\nc{\lvac}[1]{\langle{\cal O}#1|}
\nc{\rvacb}[1]{|{\cal O}_\beta #1\rangle}
\nc{\lvacb}[1]{\langle{\cal O}_\beta #1 |}
\nc{\bb}{\bar{\beta}}
\nc{\bt}{\tilde{\beta}}
\nc{\ctH}{\tilde{\cal H}}
\nc{\chH}{\hat{\cal H}}
\nc{\1}{\aa}
\nc{\2}{\"{a}}
\nc{\3}{\"{o}}
\nc{\4}{\AA}
\nc{\5}{\"{A}}
\nc{\6}{\"{O}}
\nc{\al}{\alpha}
\nc{\g}{\gamma}
\nc{\Del}{\Delta}
\nc{\e}{\epsilon}
\nc{\eps}{\epsilon}
\nc{\lam}{\lambda}
\nc{\om}{\omega}
\nc{\Om}{\Omega}
\nc{\ve}{\varepsilon}
\nc{\mn}{{\mu\nu}}
\nc{\vp}{\varphi}

\nc{\advp}[3]{{\it  Adv.\ in\ Phys.\ }{{\bf #1} {(#2)} {#3}}}
\nc{\annp}[3]{{\it  Ann.\ Phys.\ (N.Y.)\ }{{\bf #1} {(#2)} {#3}}}
\nc{\apl}[3]{{\it  Appl. Phys. Lett. }{{\bf #1} {(#2)} {#3}}}
\nc{\apj}[3]{{\it  Ap.\ J.\ }{{\bf #1} {(#2)} {#3}}}
\nc{\apjl}[3]{{\it  Ap.\ J.\ Lett.\ }{{\bf #1} {(#2)} {#3}}}
\nc{\app}[3]{{\it Astropart.\ Phys.\ }{{\bf #1} {(#2)} {#3}}}
\nc{\cmp}[3]{{\it  Comm.\ Math.\ Phys.\ }{{ \bf #1} {(#2)} {#3}}}
\nc{\cqg}[3]{{\it  Class.\ Quant.\ Grav.\ }{{\bf #1} {(#2)} {#3}}}
\nc{\epl}[3]{{\it  Europhys.\ Lett.\ }{{\bf #1} {(#2)} {#3}}}
\nc{\ijmp}[3]{{\it Int.\ J.\ Mod.\ Phys.\ }{{\bf #1} {(#2)} {#3}}}
\nc{\ijtp}[3]{{\it Int.\ J.\ Theor.\ Phys.\ }{{\bf #1} {(#2)} {#3}}}
\nc{\jmp}[3]{{\it  J.\ Math.\ Phys.\ }{{ \bf #1} {(#2)} {#3}}}
\nc{\jpa}[3]{{\it  J.\ Phys.\ A\ }{{\bf #1} {(#2)} {#3}}}
\nc{\jpc}[3]{{\it  J.\ Phys.\ C\ }{{\bf #1} {(#2)} {#3}}}
\nc{\jap}[3]{{\it J.\ Appl.\ Phys.\ }{{\bf #1} {(#2)} {#3}}}
\nc{\jpsj}[3]{{\it J.\ Phys.\ Soc.\ Japan\ }{{\bf #1} {(#2)} {#3}}}
\nc{\lmp}[3]{{\it Lett.\ Math.\ Phys.\ }{{\bf #1} {(#2)} {#3}}}
\nc{\mpl}[3]{{\it  Mod.\ Phys.\ Lett.\ }{{\bf #1} {(#2)} {#3}}}
\nc{\ncim}[3]{{\it  Nuov.\ Cim.\ }{{\bf #1} {(#2)} {#3}}}
\nc{\np}[3]{{\it  Nucl.\ Phys.\ }{{\bf #1} {(#2)} {#3}}}
\nc{\npps}[3]{{\it  Nucl.\ Phys.\ Proc.\ Suppl.\ }{{\bf #1} {(#2)} {#3}}}
\nc{\pr}[3]{{\it Phys.\ Rev.\ }{{\bf #1} {(#2)} {#3}}}
\nc{\pra}[3]{{\it  Phys.\ Rev.\ A\ }{{\bf #1} {(#2)} {#3}}}
\nc{\prb}[3]{{\it  Phys.\ Rev.\ B\ }{{{\bf #1} {(#2)} {#3}}}}
\nc{\prc}[3]{{\it  Phys.\ Rev.\ C\ }{{\bf #1} {(#2)} {#3}}}
\nc{\prd}[3]{{\it  Phys.\ Rev.\ D\ }{{\bf #1} {(#2)} {#3}}}
\nc{\prl}[3]{{\it Phys.\ Rev.\ Lett.\ }{{\bf #1} {(#2)} {#3}}}
\nc{\pl}[3]{{\it  Phys.\ Lett.\ }{{\bf #1} {(#2)} {#3}}}
\nc{\prep}[3]{{\it Phys.\ Rep.\ }{{\bf #1} {(#2)} {#3}}}
\nc{\prsl}[3]{{\it Proc.\ R.\ Soc.\ London\ }{{\bf #1} {(#2)} {#3}}}
\nc{\ptp}[3]{{\it  Prog.\ Theor.\ Phys.\ }{{\bf #1} {(#2)} {#3}}}
\nc{\ptps}[3]{{\it  Prog\ Theor.\ Phys.\ suppl.\ }{{\bf #1} {(#2)} {#3}}}
\nc{\physa}[3]{{\it  Physica\ A\ }{{\bf #1} {(#2)} {#3}}}
\nc{\physb}[3]{{\it  Physica\ B\ }{{\bf #1} {(#2)} {#3}}}
\nc{\phys}[3]{{\it Physica\ }{{\bf #1} {(#2)} {#3}}}
\nc{\rmp}[3]{{\it  Rev.\ Mod.\ Phys.\ }{{\bf #1} {(#2)} {#3}}}
\nc{\rpp}[3]{{\it Rep.\ Prog.\ Phys.\ }{{\bf #1} {(#2)} {#3}}}
\nc{\sjnp}[3]{{\it Sov.\ J.\ Nucl.\ Phys.\ }{{\bf #1} {(#2)} {#3}}}
\nc{\spjetp}[3]{{\it Sov.\ Phys.\ JETP\ }{{\bf #1} {(#2)} {#3}}}
\nc{\yf}[3]{{\it Yad.\ Fiz.\ }{{\bf #1} {(#2)} {#3}}}
\nc{\zetp}[3]{{\it Zh.\ Eksp.\ Teor.\ Fiz.\  }{{\bf #1}  {(#2)} {#3}}}
\nc{\zp}[3]{{\it Z.\ Phys.\ }{{\bf #1} {(#2)} {#3}}}
\nc{\ibid}[3]{{\sl ibid.\ }{{\bf #1} {#2} {#3}}}
\nc{\rf}[1]{(\ref{#1})}
\nc{\nn}{\nonumber \\*}
\nc{\bfB}{\bf{B}}
\nc{\bfv}{\bf{v}}
\nc{\bfx}{\bf{x}}
\nc{\bfy}{\bf{y}}
\nc{\vx}{\vec{x}}
\nc{\vy}{\vec{y}}
\nc{\oB}{\overline{B}}
\nc{\oI}{\overline{I}}
\nc{\oR}{\overline{R}}
\nc{\rar}{\rightarrow}
\nc{\ti}{\times}
\nc{\slsh}{\hskip-5pt/}
\nc{\sm}{Standard~Model~}
\nc{\MP}{M_{\rm Pl}}
\nc{\tp}{t_{\rm Pl}}
\nc{\ave}{\bar{E}}

\nc{\eff}{{\rm eff}}
\nc{\kk}{\vek{k}}
\nc{\pp}{{\rm p}}
\nc{\ga}{g_{a\gamma}}
\nc{\vv}{\\}
\nc{\eee}{{\bf E}}
\nc{\bbb}{{\bf B}}
\nc{\qcd}{T_{\rm QCD}}
\nc{\G}{\rm \ G}
\def\vec#1{{\bf #1}}
\def\lae{\;^{<}_{\sim} \;} \def\gae{\; ^{>}_{\sim} \;} 

\def\ell{e^{c}LL}

\begin{document}
{\title{\vskip-2truecm{\hfill {{\small \\
	\hfill \\
	}}\vskip 1truecm}
{Supersymmetric Curvatons and Phase-Induced Curvaton Fluctuations}}
%\vspace{1.2cm}
{\author{
{\sc John McDonald$^{1}$}\\
{\sl\small Dept. of Mathematical Sciences, University of Liverpool,
Liverpool L69 3BX, England}
}
\maketitle
%\vspace{1cm}
%\newpage
\begin{abstract}
\noindent

          We consider the curvaton scenario in the context 
of supersymmetry (SUSY) with gravity-mediated SUSY breaking. 
In the case of a large initial curvaton amplitude during inflation and
a negative order $H^{2}$ correction to the mass squared term after inflation, 
the curvaton will be close to
the minimum of its potential at the end of inflation. 
In this case the curvaton amplitude fluctuations will be
damped due to oscillations around the 
effective minimum of the curvaton potential, requiring a large
 expansion rate during inflation 
in order to account for the observed energy density perturbations,
in conflict with cosmic microwave background constraints. 
 Here we introduce a new curvaton scenario,
 the phase-induced curvaton scenario, in which de Sitter fluctuations
 of the phase of a complex SUSY curvaton field induce an 
amplitude fluctuation which is unsuppressed even 
in the presence of a negative order $H^{2}$ correction and 
large initial curvaton amplitude. This scenario is closely related to 
the Affleck-Dine mechanism and a curvaton asymmetry is naturally 
generated in conjunction with the energy density perturbations. 
Cosmological energy density perturbations can 
be explained with an expansion rate $H \approx 10^{12} \GeV$ during inflation.

\end{abstract}
\vfil
\footnoterule
{\small $^1$mcdonald@amtp.liv.ac.uk}

\thispagestyle{empty}
\newpage
\setcounter{page}{1}

\section{Introduction}

         There has recently been considerable interest in the possibility that the 
energy density perturbations leading to structure formation could originate as 
quantum fluctuations of a scalar field other than the inflaton \cite{curv,curv2,curv3}. 
This scalar field has been called
the curvaton. A number of candidates for the 
curvaton have been suggested and analyses performed \cite{clist,snu}. The 
curvaton scenario effectively decouples the inflaton energy density from the
 observed cosmic microwave background (CMB) temperature fluctuations, 
allowing for a lower energy density during inflation.
 This could be advantageous in certain inflation models such as D-term hybrid 
inflation \cite{dti}, which requires a curvaton in order to evade generation of unacceptable 
CMB fluctuations due to cosmic strings \cite{kawa}. 
It also allows inflation to be driven by a scalar potential 
which would otherwise be ruled out by the scale-dependence of its perturbation spectrum. 

              In supersymmetric (SUSY) models with 
 gravity-mediated SUSY breaking \cite{mssm,adr}, 
the curvaton is a complex field with a non-trivial scalar
 potential consisting of conventional soft SUSY
 breaking terms of the order of the weak scale,
 order $H$ SUSY breaking corrections induced by energy densities
 in the early
 Universe \cite{h2,h22,drt}, and non-renormalizable superpotential
terms suppressed by the natural mass scale of
supergravity (SUGRA), $M = M_{Pl}/\sqrt{8 \pi}$, where $M_{Pl}$ 
is the Planck mass. The resulting SUSY 
curvaton evolution will be largely determined by 
the order $H$ corrections to the curvaton potential.

   In this paper we wish to show that there is an alternative
 curvaton scenario specific to 
gravity-mediated SUSY breaking, which we call the phase-induced  
curvaton scenario.  In this scenario 
quantum fluctuations of the phase of the curvaton are
 transferred to amplitude fluctuations
 at the onset of curvaton coherent oscillations. 
This scenario will be important in 
the case where there is a large curvaton 
amplitude during inflation together with
a negative order $H^{2}$ curvaton
mass squared term after inflation.  
As we will show, in this case the conventional curvaton scenario 
based on amplitude fluctuations is typically 
inconsistent with CMB constraints.

            The possibility that the phase field of a SUSY curvaton
 could play a role in generating the energy density
 perturbations has been considered previously in the 
context of gauge-mediated SUSY breaking \cite{gmcurv}. The
 scenario discussed here 
is quite different; the phase field potential does not directly provide the 
energy density fluctuations 
(the energy density of the phase field being completely negligible in models with
 gravity-mediated SUSY breaking), but induces
 a fluctuation in the amplitude field which then serves as the curvaton as usual. 
 The final curvaton energy density perturbation in the phase-induced curvaton scenario is
 typically of the same magnitude as in the original curvaton
 scenario (which has an effectively massless curvaton up to the onset of coherent curvaton oscillations) \cite{curv3}, 
but with the value of the curvaton amplitude during inflation fixed by the 
curvaton potential. This fixes the value of $H$ during inflation in terms of the observed 
energy density perturbation and the dimension of the dominant Planck-scale suppressed 
non-renormalizable curvaton superpotential term.

     The phase-induced curvaton scenario is closely related to the
Affleck-Dine mechanism \cite{ad} and we will show that a curvaton 
asymmetry is naturally generated with a spatial perturbation 
correlated with the energy density perturbation.

          The paper is organized as follows. 
In Section 2 we discuss the curvaton
 scenario in the context of SUSY models and the evolution of the curvaton
 for the case of a large initial amplitude during inflation.  In 
Section 3 we discuss the phase-induced 
curvaton scenario. In Section 4 we present our conclusions. 

\section{Flat Direction SUSY Curvaton with a Large Initial Amplitude}

          We consider the curvaton to be a SUSY flat direction field with
 no renormalizable superpotential terms
\footnote{An example is provided by a Dirac right-handed sneutrino, although this
 requires additional interactions in order to ensure no dangerous LSP density
 from late curvaton decay \cite{snu}.}. The scalar potential then has the form 
\be{e1} V = (m_{s}^{2} + c H^{2}) |\Phi|^{2} + ( AW + h.c.)
 + \frac{\lambda^{2} d^{2} |\Phi|^{2(d-1)}}{M^{2(d-3)}}      ~,\ee 
where we consider a curvaton superpotential 
\be{e2}  W = \frac{\lambda \Phi^{d}}{M^{d-3}}        ~.\ee 
$m_{s} \approx 100 \GeV$ is the gravity-mediated soft SUSY breaking scalar mass term
and $cH^{2}$ is the order $H^{2}$ correction originating
 from non-zero F-terms due to energy densities in the early Universe \cite{h2,h22,drt}.
The magnitude of the non-renormalizable coupling $\lambda$ is naturally in the range 
from 1 to $1/d!$, the latter value being expected if the coupling arises from
 integrating out heavy fields in a complete theory \cite{kmr}. 
The A-term consists of a gravity-mediated term plus 
an order $H$ correction, $A=A_{s}+aH$, where $A_{s} \approx 100 \GeV$ \cite{drt}. 

         The values of $c$ and $a$ depend upon the couplings of the inflaton superfield 
to the curvaton \cite{drt}. After 
inflation a value of $|c|$ of the order of 1 is the most likely \cite{drt}.  
We will concentrate on the 
case with $c \approx -1$, since a positive value of $c$ will result in 
a highly damped curvaton amplitude at the onset of curvaton oscillations, making it
difficult for the curvaton to dominate the energy density before it decays.
During inflation the value of $|c|$ depends upon the inflation model. 
If inflation is driven by an F-term
then the most likely value is $|c| \approx 1$, whereas if it 
is driven by a D-term (as in D-term inflation \cite{dti}) 
then $|c| = 0$. The latter is favoured in order to have a sufficiently
 flat inflaton potential. 
The value of $a$ depends upon the coupling of the inflaton superfield to
 the curvaton. If there is no linear coupling of the inflaton to
 the curvaton in the K\"ahler potential,
 as may occur as a result of a discrete symmetry ($S \leftrightarrow -S$)
 or an R-symmetry (commonly
 introduced to eliminate dangerous non-renormalizable 
inflaton superpotential corrections 
in SUSY hybrid inflation models \cite{dti,kmr}), then $a = 
0$ throughout \cite{drt}.
 This case will be fundamental to the phase-induced curvaton scenario to be 
discussed in the next section. 

      We first review the conditions for a scalar field to play the role of a
 curvaton. Let $\Phi = \phi e^{i \theta}/\sqrt{2}$ and consider $\theta= 0$ for now. 
A basic condition for the amplitude field 
$\phi$ to serve as a curvaton is that the Universe becomes matter dominated
 by curvaton oscillations before the era of nucleosynthesis. For the case 
 $c \approx -1$ after inflation, the curvaton oscillations
 begin once $m_{s}^{2} \approx |c| H^{2}$. The largest natural 
value of $\phi$ at the onset of curvaton oscillations will correspond to
 $\phi$ approximately at the minimum of its potential when $|c|H^{2} \gae m_{s}^{2}$. 
The minimum of the potential and so initial amplitude 
can be estimated by neglecting the A-terms in the curvaton potential, since 
 the A-terms are of the same order as the other terms 
when $|c|H^{2} \approx m_{s}^{2}$.  Minimizing 
\be{e3} V(\phi) \approx \frac{c H^{2}}{2} \phi^{2}
 + \frac{\lambda^{2} d^{2} \phi^{2(d-1)} }{2^{d-1} M^{2(d-3)} }    ~,\ee
gives the minimum as a function of $H$ and $d$
 for $|c|H^{2} \gae m_{s}^{2}$,
\be{e4} \phi_{m} \approx \left( \frac{ |c| 2^{d-2}}{\lambda^{2} d^{2} (d-1)}
 \right)^{\frac{1}{2d-4}} \left(M^{2\left(d-3\right)} H^{2}
\right)^{\frac{1}{2d-4}}      ~.\ee
Coherent oscillations about $\phi = 0$ begin once $H < H_{osc} = m_{s}/|c|$. Let 
the value of $\phi$ at the onset of oscillations be $\phi_{osc}$.  
The energy density in the coherently oscillating curvaton is then
\be{e5} \rho_{\phi} \approx \left( \frac{a_{osc}}{a} \right)^{3} 
\frac{1}{2} m_{s}^{2} \phi_{osc}^{2}     ~,\ee
where $a$ is the scale factor. We assume for now that the curvaton oscillations 
begin when the Universe is matter dominated by inflaton oscillations 
(we refer to this as inflaton matter domination, IMD). 
This is true for reheating temperatures less than the thermal 
gravitino upper bound, $T_{R} \lae 10^{8} \GeV$ \cite{grav}, since
$H_{R} \lae 1 \GeV \ll H_{osc} \approx m_{s}$. 
IMD continues until the Universe becomes 
radiation dominated at the reheating temperature $T_{R}$. 
In this case  
\be{e5a} \rho_{\phi} \approx \left(\frac{a_{osc}}{a_{R}}\right)^{3} 
\left(\frac{a_{R}}{a}\right)^{3} \frac{1}{2} m_{s}^{2} \phi_{osc}^{2}  
\equiv \left(\frac{g\left(T\right)}{g\left(T_{R}\right)}\right)
 \left(\frac{T}{T_{R}}\right)^{3} 
\left(\frac{H_{R}}{H_{osc}}\right)^{2} 
\frac{1}{2} m_{s}^{2} \phi_{osc}^{2}   ~,\ee
where $g(T)$ is the number of degrees of 
freedom in thermal equilibrium and  
$H_{R}$ is the expansion rate at $T_{R}$. The energy density in the 
oscillating curvaton becomes equal to the background 
radiation density, $\rho_{rad} = \frac{\pi^{2}}{30} g(T) T^{4}$, once 
\be{e6} \phi_{osc}^{2} \approx \frac{6 M^{2}}{|c|} \left(\frac{T}{T_{R}}\right)    
 ~.\ee
With $\phi_{osc} \approx \phi_{m}(H_{osc})$, the condition that the curvaton
 dominates the energy density before curvaton decay at $T_{d}$ then requires that 
\be{e7} k_{d} \left(M^{2\left(d-3\right)} m_{s}^{2}\right)^{\frac{1}{d-2}} \gae 
\frac{6 M^{2}}{|c|} \left(\frac{T_{d}}{T_{R}}\right) \approx 
3.5 \times 10^{26} \frac{1}{|c|} \left( \frac{T_{d}}{1 \MeV}\right) 
\left(\frac{10^{8} \GeV}{T_{R}}\right) \GeV  ~,\ee
where 
\be{e8} k_{d} = \left( \frac{2^{d-2}}{\lambda^{2} d^{2} (d-1)} \right)^{\frac{1}{d-2}}
~.\ee
In this we have scaled $T_{d}$ to the temperature at nucleosynthesis, 
$T_{nuc} \approx 1 \MeV$, since the curvaton must decay before nucleosynthesis.
For $d = 4,5,6$ the left hand side of \eq{e7} is 
\be{e9a} k_{4} \left(M^{2} m_{s}^{2}\right)^{1/2} =  2.4 \times 10^{20} k_{4}
\left( \frac{m_{s}}{100 \GeV} \right)  \GeV   ~,\ee
\be{e9b} k_{5} \left(M^{4} m_{s}^{2}\right)^{1/3} =  6.9 \times 10^{25} k_{5}
\left( \frac{m_{s}}{100 \GeV} \right)^{2/3}  \GeV   ~\ee
and
\be{e9c} k_{6} \left(M^{6} m_{s}^{2}\right)^{1/4} =  3.7 \times 10^{28} k_{6}
\left( \frac{m_{s}}{100 \GeV} \right)^{1/2} \GeV   ~.\ee
The value of $k_{d}$ is approximately in the range 1 to 10 for
 $\lambda$ varying from 1 to $1/d!$.
 With $k_{d} \approx 1$ this implies that $d \geq 5$ must be satisfied 
(with $d=5$ marginal) in order to 
satisfy the curvaton condition if $T_{R}$ satisfies the conventional thermal gravitino 
upper bound on the reheating temperature, $T_{R} \lae 10^{8} \GeV$ \cite{grav}.

        It is possible that $T_{R}$ could be larger than the conventional 
thermal gravitino upper bound if the Universe is dominated by the curvaton energy 
density for a sufficiently long period \cite{gmcurv}. In this case the thermal gravitinos
 are diluted by entropy production. Once $T_{R} > T_{*}\approx  
10^{10}
 \GeV$ the curvaton oscillations will begin during radiation domination
 ($H_{R} > H_{osc} = m_{s}/|c|^{1/2}$). In this 
case it is straightforward to show 
that $T_{*}$ replaces $T_{R}$ on the right-hand side of \eq{e6}.
 Therefore the lower bound \eq{e7} can be relaxed by at most a 
factor $10^{-2}$. This allows $d=5$ to be more 
plausible but still rules out $d=4$. 

          The second fundamental condition to have a 
successful curvaton scenario 
is that the magnitude of the energy density perturbation 
should be consistent with CMB observations. This requires that  
$\delta_{\rho} = 2 \delta_{\phi} \approx 10^{-5}$, where
$\delta_{\rho} = \delta \rho/\rho$ is the energy density perturbation 
and $\delta_{\phi} = \delta \phi/\phi$ is the perturbation 
of the curvaton oscillation amplitude ($\rho \propto \phi^{2}$). 
The value of $\delta_{\phi}$ depends upon the 
evolution of the curvaton after inflation.  
For $c \approx -1$ after inflation there are two possibilities:
\newline (i) $\phi$, initially small, is still evolving 
towards $\phi_{m}$ at the onset of coherent oscillations. 
(This possibility was discussed in \cite{snu}.)
\newline (ii) $\phi$ reaches $\phi_{m}$ before the onset of oscillations.

         In case (ii) the curvaton amplitude perturbation at the onset of oscillations 
will depend upon when $\phi$ reaches $\phi_{m}$. We will consider the
 case where $\phi$ is close to $\phi_{m}$ at the end of inflation. 
This will be the
 situation if $c \approx -1$ during inflation, as in F-term inflation, in which 
case the curvaton amplitude will be rapidly damped to $\phi_{m}(H_{I})$. 
It will also be true in the case where $c \approx 0$ during
 inflation, as in D-term inflation, if at the onset of inflation 
 the flat direction field has a value
 greater than the value of the minimum at the end of inflation, 
$\phi_{m}(H_{I})$. For example, this would be expected if the Universe evolves from
chaotic initial conditions such that $V(\phi) \approx M^{4}$ initially \cite{linci,jc}. 
It would also be expected if all values of $\phi$ at the onset of inflation
 were equally likely up to the value at which the 
curvaton effective mass becomes dynamically significant,  
$V^{''}(\phi) \approx H_{I}^{2}$. In this case the average value of $\phi$ at the
onset of inflation will satisfy $\phi \gae \phi_{*}$,
where $\phi_{*}$ denotes the value of $\phi$ at which $V^{''}(\phi) \approx H_{I}^{2}$.
(A similar argument in the context of axion cosmology has been given in \cite{linax}.) 
If initially $\phi \gae \phi_{*}$ then the curvaton will undergo rapid
 damped oscillations in a $\phi^{2(d-1)}$ potential until the amplitude of oscillation
 reaches $\phi_{*}$, shortly after which the amplitude evolution will become 
highly damped and its value effectively frozen. $\phi_{*}$ is given by 
\be{e10} \phi_{*} = \left( \frac{1}{|c| \left(2d-3\right)}\right)^{\frac{1}{2d-4}}
 \phi_{m}(H_{I})  
 ~.\ee 
Thus if the initial curvaton 
amplitude is greater than or of the order of  $\phi_{*}$
we expect $\phi \approx \phi_{*} = O(1) \phi_{m}(H_{I})$ at the end of inflation. 
In the following we will assume that $\phi$ 
reaches $\phi_{*}$ before the length scales
relevant to cosmological structure formation leave the horizon, since otherwise a 
highly scale-dependent curvaton perturbation spectrum would be obtained \cite{jc}. 
 
        We next consider the evolution of a superhorizon spatial perturbation of the
curvaton amplitude. Assuming the curvaton is effectively massless during inflation, the
magnitude of the perturbation on a given scale at horizon crossing is $\delta \phi =
H_{I}/2 \pi$. The scalar field equation is
\be{e11} \ddot{\phi} + 3 H \dot{\phi} 
- \frac{\vec{\nabla}^{2}}{a^2}\phi  = - V^{'}\left(\phi \right)      ,~\ee
where $\phi(\vec{x},t) = \phi_{o}(t) + \delta \phi(\vec{x},t)$. $\phi_{o}(t)$ is the 
homogeneous curvaton amplitude and $\delta  \phi(\vec{x},t)$ is the 
spatial fluctuation due to quantum fluctuations. For a fluctuation of wave number $k 
\ll H$, the gradient term in \eq{e11} is negligible and the evolution of the field at a
 point in space is determined by  
\be{e12} \ddot{\phi} + 3 H \dot{\phi} \approx - V^{'}\left(\phi \right)      .~\ee
For $|c|H^{2} \gg m_{s}^{2}$ and with $H \propto a^{-m}$, 
a solution of \eq{e12} 
is given by $\phi =\overline{\phi}_{m} \equiv K \phi_{m}$, where  
\be{e13} K^{2d-4} = 1 + \frac{1}{|c| \left(d-2\right)} \left(3m - m^{2}
 \left(\frac{d-1}{d-2}\right) \right)     ~.\ee  
If the curvaton amplitude is initially at or close to $\overline{\phi}_{m}$ at the end of 
inflation, then the subsequent evolution of $\phi$ at a point in space 
will correspond to oscillations about $\overline{\phi}_{m}$. 
Substituting $\phi = \overline{\phi}_{m} + \Delta \phi$, where $\Delta \phi (\vec{x},t) = 
\Delta \phi_{o}(t) + \delta \phi(\vec{x},t) \ll 
\overline{\phi}_{m}$ and $\Delta \phi_{o}(t) = \phi_{o} - \overline{\phi}_{m}$, 
the equation for a superhorizon perturbation $\delta \phi (\vec{x},t)$ 
at a point in space becomes
\be{e14} \delta \ddot{\phi} + 3 H \delta \dot{\phi}  \approx  - \alpha H^{2} \delta \phi  
 ~,\ee
 where 
\be{e15} \alpha = |c| \left(\left(2d-3\right) K^{2d-4} - 1 \right)   ~.\ee
($\Delta \phi_{o}(t)$ satisfies the same equation.) The general solution of \eq{e14} 
 is $\delta \phi \propto a^{\gamma}$,  where 
\be{e16} \gamma = \frac{1}{2} \left[ - \left(3-m\right) + \sqrt{\left(3-m\right)^{2}
 - 4 \alpha}  \right]    ~.\ee
Thus, with $\delta \phi(\vec{x},t) = \delta \phi(t) \sin(\vec{k}.\vec{x})$, 
the evolution of the amplitude of a superhorizon 
quantum fluctuation at a point in space is given by $\delta \phi(t) \propto a^{\gamma}$. 
We can then compute the suppression of $\delta \phi$ due to expansion 
from the end of inflation until the
 onset of coherent curvaton oscillations. We assume that coherent curvaton 
oscillations begin during IMD, such that $m = 3/2$. Then 
$\gamma = \frac{1}{2} \left[ - \frac{3}{2} + \sqrt{ \frac{9}{4} - 4 \alpha} \right]$. 
Since $\alpha \sim 1$, we expect an oscillating solution for $\delta \phi (t)$
 (i.e. imaginary root in $\gamma$) with oscillation amplitude scaling as $\delta \phi \propto
 a^{-3/4}$.  

If $\phi_{o}$ is initially close to $\overline{\phi}_{m}$ 
then since $\Delta \phi_{o}/\overline{\phi}_{m}$ decreases with time, 
the mean value of the 
curvaton amplitude will become increasingly close to the effective minimum, 
 $\phi_{o} \approx \overline{\phi}_{m} \propto 
H^{\frac{1}{d-2}} \propto a^{-\frac{3}{2\left(d-2\right)}}$. 
 Thus if $\phi_{o} \approx \overline{\phi}_{m}$ at the end of inflation
 (corresponding to scale factor $a_{e}$) 
we have
\be{e17}  \left(\frac{\delta \phi}{\phi}\right)_{osc} \approx 
 \left(\frac{a_{e}}{a_{osc}}\right)^{\frac{3}{4}\left(1 
- \frac{2}{\left(d-2\right)}\right)} \left(\frac{\delta \phi}{\phi}\right)_{I}  ~,\ee
where $(\delta \phi/\phi)_{I}$ is the value of the quantum fluctuation 
at the end of inflation. ($(\delta \phi/\phi)_{I}$ is constant 
during inflation since $\phi$ is frozen for $\phi \lae \phi_{*}$.)
Once curvaton coherent oscillations begin, $\delta_{\phi}$ remains constant since 
$V(\phi) \propto \phi^{2}$ and both $\phi_{o}$ and $\delta \phi$ evolve 
in the same way ($\propto a^{-3/2}$).  
Since $K$ is very close to 1 ($K = 1.044$ for $d=6$, $m=3/2$ and $|c| = 1$)
we can assume $\overline{\phi}_{m} \approx \phi_{m}$.  
With $\phi \approx \phi_{m}(H_{I})$ and $\delta \phi \approx 
H_{I}/2\pi$ when the curvaton perturbations leave the horizon, we obtain
\be{e18} \delta_{\rho}  \approx  2 \left(\frac{\delta \phi}{\phi}\right)_{osc} \approx 
\left(\frac{H_{osc}}{H_{I}}\right)^{\frac{1}{2}\left(1 -
 \frac{2}{d-2}\right)} \frac{H_{I}}{\pi \phi_{m}\left(H_{I}\right)}      ~.\ee
This fixes $H_{I}$ in terms of $\delta_{\rho}$ and $d$. With 
$\phi_{m}(H_{I}) \approx k_{d}^{1/2} M^{\frac{d-3}{d-2}} H_{I}^{\frac{1}{d-2}}$, the
 expansion rate during inflation is given by 
\be{e19} H_{I} \approx  \pi^{2} k_{d} M^{\frac{2\left(d-3\right)}{d-2}} 
\left( \frac{|c|^{1/2}}{m_{s}} \right)^{\frac{d-4}{d-2}} \delta_{\rho}^{2}     ~.\ee
For example, for $d=6$ this requires that 
\be{e20}  H_{I} \approx \frac{\pi^{2} k_{6} |c|^{1/4} \delta_{\rho}^{2} 
 M^{3/2}}{m_{s}^{1/2}} \approx 3.7 \times 10^{17} |c|^{1/4}
k_{6} \left(\frac{\delta_{\rho}}{10^{-5}}\right)^{2} \left(\frac{100
 \GeV}{m_{s}}\right)^{1/2}  \GeV     ~,\ee
whilst for $d=5$
\be{e20a}  H_{I} \approx \frac{\pi^{2} k_{5} |c|^{1/6} \delta_{\rho}^{2} 
 M^{4/3}}{m_{s}^{1/3}} \approx 6.8 \times 10^{14} |c|^{1/6}
k_{5} \left(\frac{\delta_{\rho}}{10^{-5}}\right)^{2} \left(\frac{100
 \GeV}{m_{s}}\right)^{1/3}  \GeV     ~,\ee
where $k_{5}$, $k_{6} \gae 1$. These values of $H_{I}$ correspond to an energy density which would
result in unacceptably large energy density perturbations from conventional inflaton
 quantum fluctuations, which typically require $H_{I} \lae 10^{13} \GeV$ in order 
to be compatible with CMB constraints. Thus 
in the case of a large curvaton amplitude during inflation and negative order 
$H^{2}$ correction after inflation, the energy 
density perturbations due to curvaton amplitude 
fluctuations are typically inconsistent with CMB constraints.

          In the above we have assumed that quantum de Sitter fluctuations
 of the curvaton field during inflation are unsuppressed, i.e. that the curvaton effective
 mass is much smaller than $H_{I}$. 
In the case of inflation driven by an F-term we expect
 that $c \approx -1$ for the flat direction field
 during inflation. In this case there will be effectively no 
superhorizon curvaton amplitude fluctuations on the scale of
 large scale structure formation,
 as these will be highly damped by curvaton evolution
 due to the order $H^{2}$ term during inflation. Thus a curvaton scenario based on 
fluctuations of the curvaton amplitude is ruled out in this case also. 

     However, as we will show in the next 
section, in both the large initial curvaton amplitude 
and F-term inflation cases it is still possible for a complex SUSY curvaton with 
gravity-mediated SUSY breaking to generate an energy density perturbation which is  
consistent with CMB constraints, via fluctuations of the curvaton phase.

\section{Phase-Induced Curvaton Scenario}

                     If the inflaton has no linear coupling to the curvaton in 
the K\"ahler potential then there will be no order $H$ correction to 
the A-terms \cite{drt,gmcurv,kmr}. This allows for a new version of 
the curvaton scenario which applies 
specifically to a complex SUSY curvaton with gravity-mediated SUSY breaking. 

            In the absence of order $H$ corrections to the A-terms, the phase field of the 
curvaton is effectively massless during and after inflation, even in the case of F-term inflation 
where the amplitude field typically gains an effective mass of order $H$. Thus de Sitter fluctuations 
of the phase will be unsuppressed. For $\phi$ constant 
during inflation, the canonically normalized phase field for
small $\delta \theta$ about an amplitude in the $\theta$ direction 
is $\phi_{p} \approx \phi_{I} \delta \theta$, where $\phi_{I}$
 is the curvaton amplitude during inflation.  
Thus with $\delta \phi_{p} = H_{I}/2\pi$ the de Sitter 
fluctuation of the phase field is $\delta \theta \approx H_{I}/2 \pi \phi_{I}$. 

     The equation of motion of the complex curvaton field is 
$$ \ddot{\Phi} + 3 H \dot{\Phi} - \frac{\vec{\nabla}^{2}}{a^{2}}\Phi 
=  - \frac{\partial V}{\partial \Phi^{\dagger}} $$
\be{e21}  \equiv - \frac{e^{i \theta} \phi}{\sqrt{2}}
 \left[ 
\left(m_{s}^{2} + c H^{2}\right) + \frac{A \lambda d \phi^{d-2}}{M^{d-3}
 \left( \sqrt{2} \right)^{d-2}} \left(\cos\left(d \theta\right) 
- i \sin \left( d \theta \right) \right) 
   + \frac{ \lambda^{2} d^{2} \left(d-1\right) \phi^{2 \left(d-2\right)} }{2^{d-2}
 M^{2 \left(d-3\right)} }  \right]   
~.\ee
For perturbations of wavelength much larger than the horizon 
the gradient terms are negligible. In this case the evolution of the curvaton
 amplitude and phase at a point in space is given by
\be{e21a}  \ddot{\phi} + 3 H \dot{\phi} - \dot{\theta}^{2} \phi = 
- \left[ 
\left(m_{s}^{2} + c H^{2} \right) \phi + \frac{A \lambda d \phi^{d-1}
 \cos\left(d \theta\right) }{\left(\sqrt{2}\right)^{d-2} M^{d-3}} 
 + \frac{\left(d-1\right) \lambda^{2} d^{2} \phi^{2d-3}}{2^{d-2} M^{2\left(d-3\right)}}
\right]  
 ~\ee
and
\be{e22} \ddot{\theta} + 3 H \dot{\theta} + \frac{2\dot{\phi}}{\phi}\dot{\theta} 
=  \frac{A \lambda d \phi^{d-2} \sin\left(d\theta\right)
 }{\left(\sqrt{2}\right)^{d-2} M^{d-3}}   ~.\ee
From \eq{e22} we see that without an A-term the phase at a point 
in space will remain constant. In the absence of order $H$ corrections to the A-term 
the phase will begin to evolve only when $m_{s}^{2} \approx |c| H^{2}$, when 
 the A-term contribution 
to the potential becomes comparable with that of the mass squared term. A
 perturbation in the phase field will then induce a perturbation in the curvaton
 amplitude via the coupling proportional to $\cos(d\theta)$ in \eq{e21a}.

\begin{figure}[hp]
\begin{center}
\includegraphics[width=0.75\textwidth]{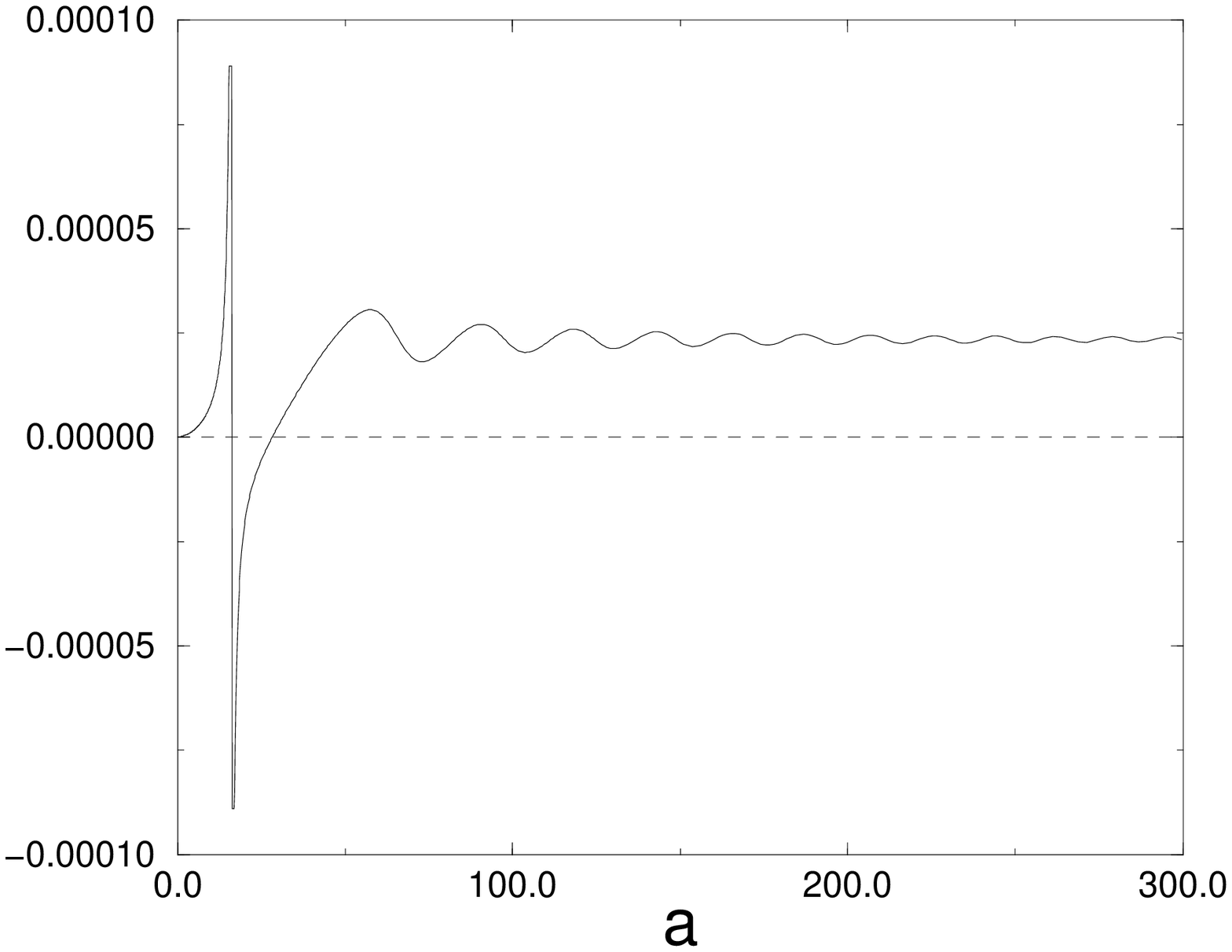}
\caption{\footnotesize{Evolution of the energy density perturbation for $d=6$ and $A/m_{s} = 1$.}}
\label{fig:fig1}
\end{center}
%\end{figure}

%\begin{figure}[hp]
\begin{center}
\includegraphics[width=0.75\textwidth]{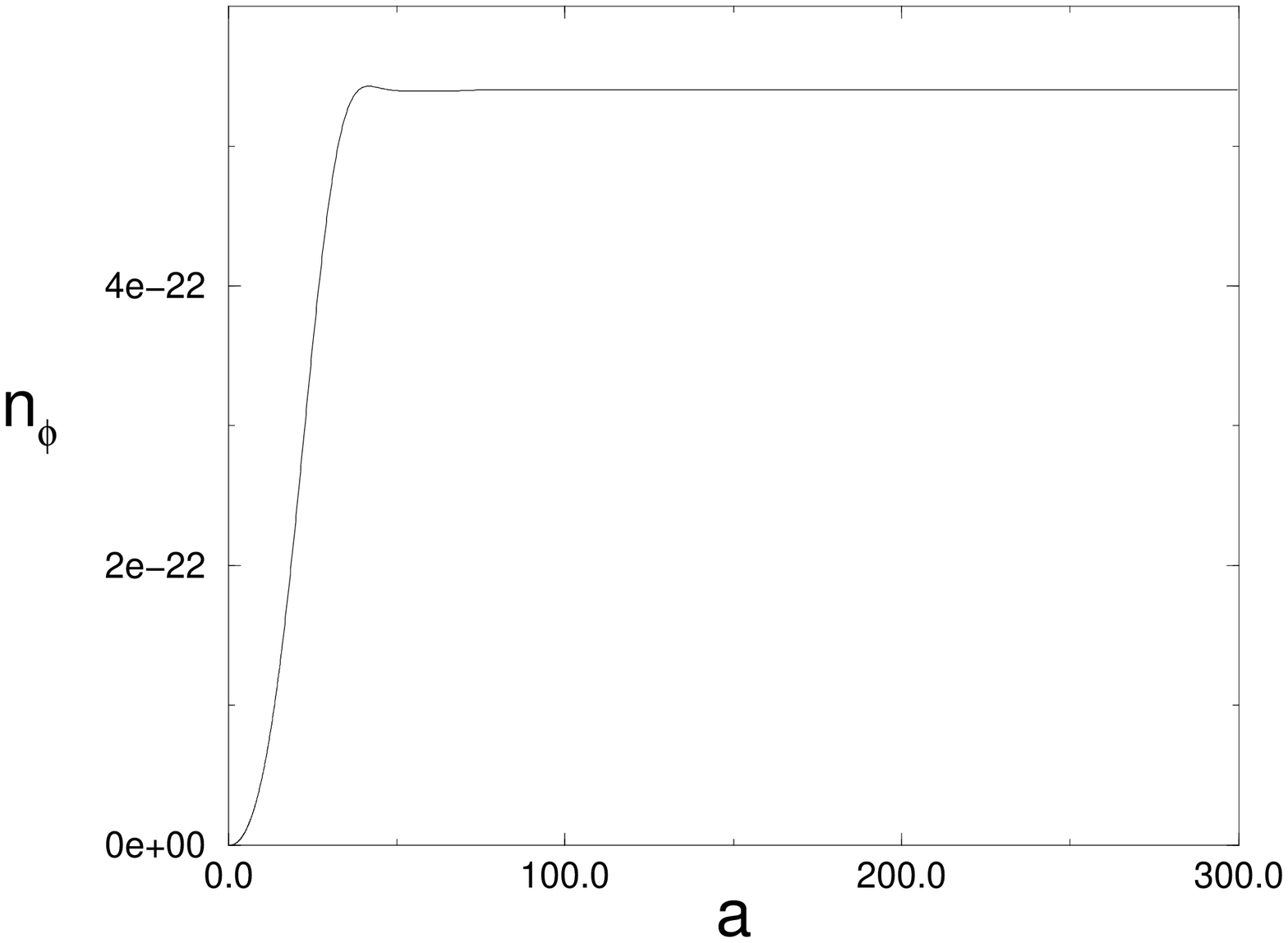}
\caption{\footnotesize{Evolution of the comoving curvaton asymmetry for $d=6$ and $A/m_{s} = 1$.}}
\label{fig:fig2}
\end{center}
\end{figure}

\begin{figure}[hp]
\begin{center}
\includegraphics[width=0.75\textwidth]{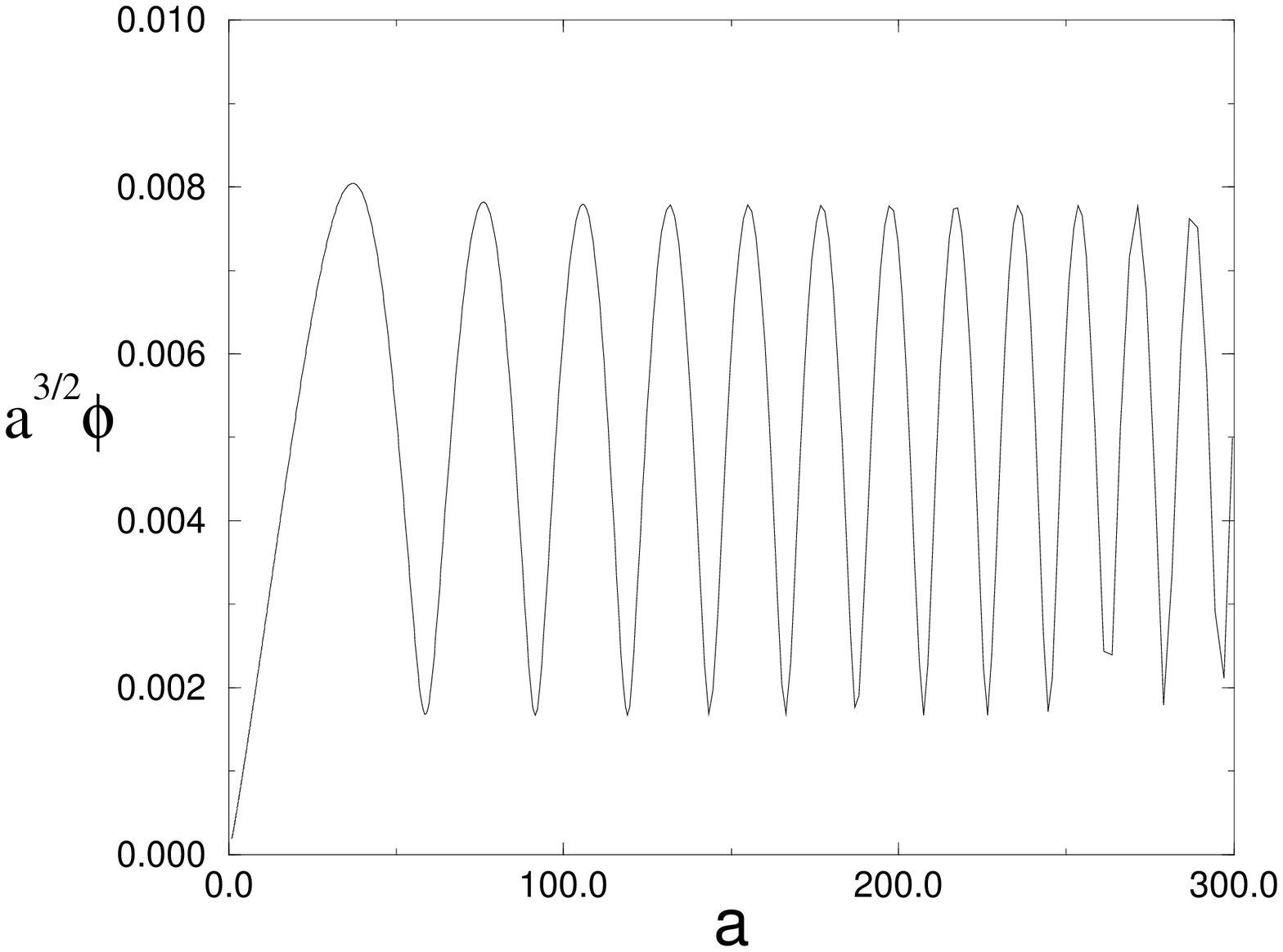}
\caption{\footnotesize{Evolution of the curvaton amplitude for $d=6$ and $A/m_{s} = 1$.}}
\label{fig:fig3}
\end{center}
%\end{figure}

%\begin{figure}[h]
\begin{center}
\includegraphics[width=0.75\textwidth]{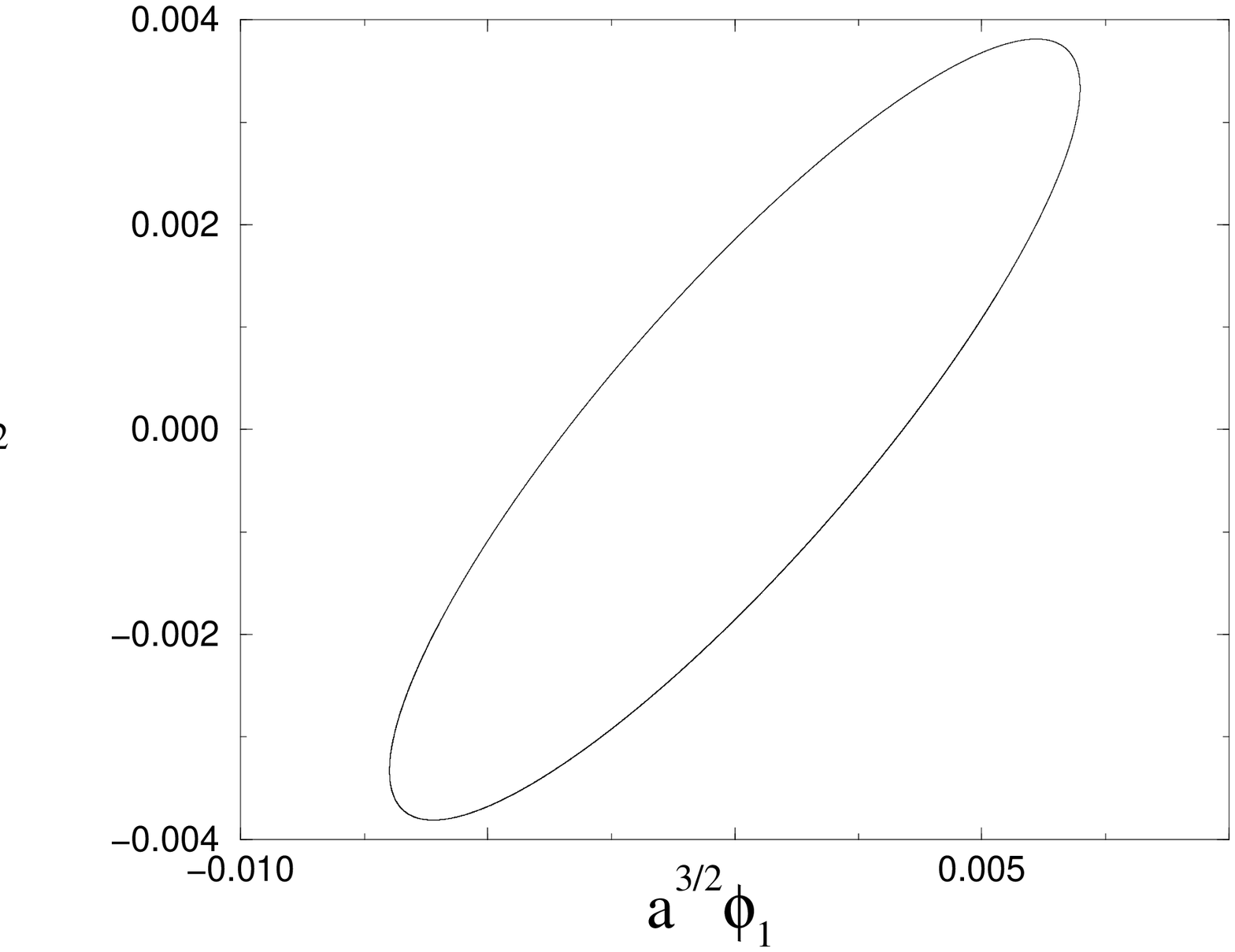}
\caption{\footnotesize{Late-time trajectory of the curvaton in the complex plane.}}
\label{fig:fig4}
\end{center}
\end{figure}

    In our numerical results we focus on the case $d = 6$ as an example. We assume that 
$c = -1$ and $\lambda = 1$ throughout. (We use units such that $M = 1$.)
We evolve the equations of motion from a time corresponding to initial
 expansion rate $H^{2} = 8000 m_{s}^{2} \gg |c|H^{2}$, 
with initial values $\phi = \overline{\phi}_{m}$, 
$\dot{\phi} = \dot{\overline{\phi}}_{m}$, $d \theta = \pi/4$ and
 $\dot{\theta} = 0$. We set $a = 1$ at this initial time. For the perturbations we assume
 $\delta \phi = 0$ and $\delta \theta = 10^{-5}$ initially. To calculate the evolution of the
 perturbation we evolve \eq{e21a} and \eq{e22} for two points in space, one with initial 
value $\theta$ and one with initial value 
$\theta + \delta \theta$, corresponding to points in space with 
the mean value of $\theta$ and the largest perturbed value of $\theta$   
respectively.

               In Figure 1 we show the evolution of the energy density perturbation 
with scale factor for the case $A/m_{s} = 1$. 
The spike feature is associated with the 
vanishing of the curvaton mass squared term at scale
 factor $a_{osc}$ ($= 19.44$ in our units) corresponding to $|c|H^{2} = m_{s}^{2}$. 
At around this time the energy density briefly 
vanishes whilst changing sign, resulting in a large value for $\delta \rho/\rho$.   
Once $a$ is large compared with $a_{osc}$, 
the non-renormalizable terms in the curvaton potential become 
negligible as a result of the reduction of the curvaton amplitude due to
expansion. 
Curvaton evolution in a $|\Phi|^{2}$ potential is then established with a constant 
energy density perturbation, numerically given by $\delta_{\rho} = 2.3 \times 10^{-5}$ 
for the case $A/m_{s} = 1$ and $\delta \theta = 10^{-5}$. 

            The phase-induced curvaton scenario is closely related to the 
Affleck-Dine (AD) mechanism for the  
generation of global charge asymmetries such as baryon number \cite{ad}. 
The explicit dependence of the A-term on the phase of the scalar field introduces
 CP-violation into the equations of motion. The result is that the real and imaginary
 parts of $\Phi$ coherently oscillate out of phase with each other in a 
$|\Phi|^{2}$ potential once $a \gg a_{osc}$, such that the
 complex curvaton field describes an ellipse in the complex plane.
 Once $a > a_{osc}$ and the non-renormalizable corrections become negligible, 
the curvaton potential has an approximately conserved global $U(1)$ symmetry, with
 respect to which a constant curvaton asymmetry in a comoving volume is
 established. 
The comoving curvaton charge asymmetry density is given by 
\be{ad1}  n_{\phi} = a^{3} i ( \dot{\Phi}^{\dagger} \Phi - \Phi^{\dagger} \dot{\Phi} ) \equiv 
a^{3} \dot{\theta} \phi^{2}      ~,\ee
where the factor $a^{3}$ compensates for the expansion of the Universe, 
such that $n_{\phi}$ 
is constant for a conserved charge. 
In Figure 2 we show the growth of the comoving curvaton asymmetry for the case 
$A/m_{s} = 1$. 
This shows that a curvaton charge asymmetry naturally arises in 
conjunction with the energy density perturbations 
in the phase-induced curvaton scenario. $\theta$ introduces
CP violation resulting in a curvaton asymmetry whilst $\delta \theta$ 
induces the energy density perturbations. 

              In Figure 3 we show the evolution of the 
curvaton amplitude field in the form $a^{3/2} \phi$, where $a^{3/2}$ cancels out the
 effect of expansion on evolution in a $|\Phi|^{2}$ potential.
 For $a \gg a_{osc}$ we see that the curvaton 
amplitude has a maximum value $\approx 0.008$ and a minimum value $\approx 0.002$,
 corresponding to the major and minor axis of the ellipse described by the globally charged
 curvaton field in the complex plane. In Figure 4 we show the late time trajectory of the curvaton in the complex plane, where $\Phi = (\phi_{1} + i \phi_{2})/\sqrt{2}$.    

          In Table 1 we give the asymptotic values  
of $\delta_{\rho}$, $n_{\phi}$ and $\delta n_{\phi}/n_{\phi}$ 
(the spatial perturbation of the curvaton asymmetry) at $a\gg a_{osc}$ 
for different values of $A/m_{s}$. ($n_{\phi}$ gives the charge asymmetry at $a = 300$.) From this
 we see that the values of these quantities are sensitive to 
$A/m_{s}$. For $|A/m_{s}| = O(1)$ we broadly find that $|\delta_{\rho}|
 = O(1) |\delta_{\theta}|$ for $A > 0$ and 
$|\delta_{\rho}| = O(0.1) |\delta_{\theta}|$ for $A < 0$. 
The results in Table 1 are for the case $\lambda = 1$.
We find that the values of $\delta_{\rho}$ and $\delta n_{\phi}/n_{\phi}$
 are independent of the non-renormalizable coupling $\lambda$.  
The perturbation of the curvaton charge asymmetry is particularly sensitive to the value of
 $A$. In particular, the curvaton asymmetry changes sign as $A$ is 
varied, implying that the mean value of the 
asymmetry vanishes for particular values of $A$. Close to these values 
$n_{\phi}$ becomes small, implying that $\delta n_{\phi}/n_{\phi}$ becomes large.
For example, this occurs when $A/m_{s} = 2$. 
For most values of $A$ the curvaton asymmetry perturbation
 has a larger magnitude than the energy density perturbation. 
However, for some values of $A$, for example $A/m_{s} = 1$, it is much
 smaller than the 
energy density perturbation. This could be 
significant if the curvaton asymmetry was the
 source of a cosmological  baryon or lepton asymmetry, since in this case there would be a
 perturbation of the asymmetry correlated with the energy density perturbation. 
It is unlikely that the curvaton asymmetry could be directly interpreted as a baryon asymmetry, 
since this would imply that the baryon number to entropy ratio 
at curvaton decay ($n/s \approx T_{d}/m_{s} \gae 10^{-5}$ for a typical curvaton condensate with 
energy per unit charge\footnote{The energy per unit charge is equal to 
the curvaton mass $m_{s}$ when the trajectory in the complex plane is 
circular, and is of the order of $m_{s}$ 
when the trajectory is elliptical with major and minor axes of the same order of magnitude.} of the order of $m_{s}$) is 
much larger that the observed value of $10^{-10}$. 
However, a large lepton asymmetry is a possibility. It is also possible 
that a small baryon asymmetry could be generated via suppressed 
sphaleron conversion of a small fraction of this large lepton asymmetry \cite{lepta}. 
Alternatively, the curvaton asymmetry might effectively serve as a source of 
CP violation to produce a baryon asymmetry indirectly.

         From our numerical results we conclude that perturbations of the phase of the
 complex curvaton field can induce an energy density
 perturbation 
$\delta_{\rho} = f_{\theta} \delta \theta$ with $|f_{\theta}|$ of the order of 1. 
This mechanism requires only that the A-terms receive no order
 $H$ corrections during or after inflation. 

      We note that the curvaton phase field itself does not significantly
 contribute to the energy density  
once $a \gg a_{osc}$. The A-term in the scalar
 potential, responsible for the potential energy of the curvaton phase
 field, is proportional to $\phi^{d}$. 
It is therefore completely negligible compared with the
 mass squared term once $a \gg a_{osc}$.
 For this reason it is essential that in gravity-mediated SUSY breaking 
the fluctuation of the curvaton phase is transferred to a fluctuation 
of the curvaton amplitude, in order to generate a significant energy
 density perturbation once 
$a \gg a_{osc}$. This is in contrast with the case of gauge-mediated SUSY breaking, 
where it is possible for the phase field potential to directly
 provide the curvaton fluctuation \cite{gmcurv}.

\begin{table}[h]
\begin{center}
\begin{tabular}{|c|c|c|c|}
	\hline  $A/m_{s}$ & $\delta_{\rho}$ & $n_{\phi}$ & $\delta n_{\phi}/n_{\phi}$ \\
	\hline  $3.0$ & $-1.4\times 10^{-5}$ & $-1.9\times 10^{-22}$ & $-5.0\times 10^{-4}$ \\
		$2.5$ & $3.2\times 10^{-6}$ & $-7.1\times 10^{-22}$ & $-3.2\times 10^{-6}$  \\
		$2.0$ & $1.7\times 10^{-5}$ & $2.6\times 10^{-24}$ & $-3.5\times 10^{-2}$  \\
		$1.5$ & $2.6\times 10^{-5}$ & $5.3\times 10^{-22}$ & $-8.7\times 10^{-5}$  \\
		$1.0$ & $2.3\times 10^{-5}$ & $5.4\times 10^{-22}$ & $6.0\times 10^{-7}$  \\
		$0.5$ & $9.7\times 10^{-6}$ & $2.8\times 10^{-22}$ & $3.8\times 10^{-5}$  \\
		$-1$ & $1.1\times 10^{-6}$ & $-1.3\times 10^{-22}$ & $9.0\times 10^{-5}$ \\
		$-2$ & $5.5\times 10^{-7}$ & $1.6\times 10^{-22}$ & $7.5\times 10^{-5}$  \\
		$-3$ & $-1.8\times 10^{-6}$ & $-1.5\times 10^{-22}$ & $6.1 \times 10^{-5}$ \\

	\hline     

\end{tabular}
\caption{Asymptotic values of $\delta_{\rho}$, $n_{\phi}$ and 
$\delta n_{\phi}/n_{\phi}$ as a function of $A$ for the case $d=6$ and 
$\delta \theta = 10^{-5}$.}

\end{center}
\end{table}

          The energy density perturbation from the 
phase-induced curvaton scenario 
is of the same order of magnitude as that obtained in the  
original curvaton scenario \cite{curv3}
 for the case $\phi \approx 
\phi_{m}(H_{I})$, 
\be{ex1} \delta_{\rho} = f_{\theta} \delta \theta \approx  
\frac{f_{\theta} H_{I}}{2 \pi \phi_{m}\left(H_{I}\right)}    ~.\ee
In the phase-induced curvaton scenario the value of $H_{I}$ is fixed by
$\delta_{\rho}$ and $d$, 
since $\phi_{m}(H_{I})$ is determined by $H_{I}$ and $d$. This is in contrast with 
the original curvaton scenario, where the 
value of $\phi$ during inflation can take any value \cite{curv3}.
The expansion rate in the phase-induced curvaton scenario is then given by 
\be{ex2}  H_{I} \approx \left(\frac{2 \pi k_{d}^{1/2} 
\delta_{\rho}}{f_{\theta}}\right)^{\frac{d-2}{d-3}}
 M    ~.\ee
An important feature of this expression is that $H_{I}$ has
 an upper bound, corresponding to $d \rightarrow \infty$, 
\be{ex3} H_{I} \lae \frac{2 \pi  
\delta_{\rho}M}{f_{\theta}} \approx 1.5 \times 10^{14}
 \frac{1}{f_{\theta}} \left( \frac{\delta_{\rho}}{10^{-5}}\right) \GeV   ~,\ee
where we have used $k_{\infty} = 1$. 
For smaller $d$ the value of $H_{I}$ is smaller. For example, for $d = 5$ 
we obtain 
\be{ex4}  H_{I} = \left(\frac{2 \pi k_{5}^{1/2} \delta_{\rho}}{f_{\theta}}\right)^{\frac{3}{2}} M   
 \approx  1.2 \times 10^{12}  \left(\frac{k_{5}^{1/2}}{f_{\theta}}\right)^{3/2}  
\left(\frac{\delta_{\rho}}{10^{-5}}\right)^{3/2}  \GeV  ~,\ee
whilst for $d = 6$
\be{ex4a}  H_{I} = \left(\frac{2 \pi k_{6}^{1/2} \delta_{\rho}}{f_{\theta}}\right)^{\frac{4}{3}} M   
 \approx  6.0 \times 10^{12}  \left(\frac{k_{6}^{1/2}}{f_{\theta}}\right)^{4/3}  
\left(\frac{\delta_{\rho}}{10^{-5}}\right)^{4/3}  \GeV  ~.\ee
For the $d=6$ example of Table 1 we have $f_{\theta} = 2.3$ for $A = 
m_{s}$, which 
implies that $H_{I} \approx 2 \times 10^{12} \GeV$ for $k_{6} \approx 1$. 

              Thus the values of $H_{I}$ in the phase-induced curvaton scenario are 
typically less than or of the order of the values which are 
obtained in inflation models where the energy density perturbations are explained by
conventional adiabatic inflaton fluctuations.
 Therefore a complex SUSY curvaton can consistently
account for the observed energy density perturbations  
in the case of a large initial curvaton amplitude
 or F-term inflation, where 
the conventional SUSY curvaton scenario with amplitude 
fluctuations would be ruled out by CMB constraints.

\section{Conclusions}

                     We have shown that it is possible for energy density perturbations in the 
SUSY curvaton scenario with gravity-mediated SUSY breaking 
to originate from quantum fluctuations of the phase of the complex curvaton field. 
The phase fluctuations induce curvaton amplitude fluctuations 
through the SUSY breaking A-terms in the curvaton scalar potential. This requires that 
there are no order $H$ corrections to the A-term during and after inflation, which is true if
 the inflaton has no linear couplings in the K\"ahler potential. This can easily occur via a 
discrete symmetry or R-symmetry, as is necessary in SUSY hybrid inflation models.

The phase-induced curvaton scenario 
 becomes important when (i) the initial amplitude of 
the curvaton during inflation is large and 
(ii) after inflation there is a negative order $H^{2}$ correction to the curvaton mass 
squared term. Both of these are natural possibilities. In this case  
 the conventional curvaton scenario based on amplitude fluctuations 
is likely to be inconsistent with cosmic microwave background constraints. 
The phase-induced curvaton scenario, on the other hand, can consistently 
account for the observed cosmological energy density perturbations even if 
(i) and (ii) are satisfied.

    The phase-induced curvaton scenario is closely related to the Affleck-Dine mechanism 
and a curvaton asymmetry naturally occurs in conjunction with
the energy density perturbations. As a result, there are perturbations
in the curvaton asymmetry correlated with the energy density perturbations, which could 
be significant if the curvaton asymmetry was the origin of a cosmological 
baryon or lepton asymmetry. 
It would be interesting to explore further the possible connection between this version of the
curvaton scenario and particle asymmetries in cosmology.

\end{document}